# Mapping the potential within a nanoscale undoped GaAs region using a scanning electron microscope

B. Kaestner

Microelectronics Research Centre, Cavendish Laboratory, University of Cambridge, Madingley Road, Cambridge CB3 0HE, UK

C. Schönjahn, C.J. Humphreys

Department of Materials Science & Metallurgy, University of Cambridge, Pembroke Street, Cambridge CB2 3QZ, UK

**Abstract**

Semiconductor dopant profiling using secondary electron imaging in a scanning electron microscope (SEM) has been developed in recent years. In this paper, we show that the mechanism behind it also allows mapping of the electric potential of undoped regions. By using an unbiased GaAs/AlGaAs heterostructure, this article demonstrates the direct observation of the electrostatic potential variation inside a 90nm wide undoped GaAs channel surrounded by ionized dopants. The secondary electron emission intensities are compared with two-dimensional numerical solutions of the electric potential.

Imaging surface topography is widely regarded as the main application of secondary electron (SE) imaging in the scanning electron microscope (SEM). However, the SE emission intensity is also sensitive to other materials properties, e.g. work function, stopping power and electron mean free path. When SE imaging is used for semiconductors, the *p*-type region is brighter than the *n*-type region of a *p-n* junction. The main mechanism for this effect is believed to be due to a difference in the escape barrier height $E_b$ [1-3]. This difference results from matching the two Fermi levels for the *p*- and *n*-type regions. This leads to a built-in potential such that the energy bands in the *p*-type region are shifted up relative to those in the *n*-type region, thus reducing $E_b$, the energy required to escape to the detector.

This effect has been used mainly to image doped regions in semiconductors [4-9] and the technique is often called "doping contrast". In addition, the electrostatic potential distribution of a cross-section of a *p-n* junction has been mapped by subtracting two images obtained at different bias voltages [10]. The purpose of this paper is to demonstrate that via the same contrast mechanism SE imaging can be used to map with nanoscale spatial resolution the electrostatic potential in a region that is undoped. The potential variation is caused by the spatial variation of the doping concentration outside this region. In this way a gradient of the electrostatic potential is generated both vertically and laterally in the undoped region. The technique demonstrated in this paper should be useful for mapping with high spatial resolution the potential variations in a wide range of both doped and undoped nanostructures, made from semiconductors, metals, ceramics, and combinations of these materials.

The GaAs/AlGaAs heterostructure used is a modified version of a quasi-lateral nanoscale diode [11]. The starting material was prepared using molecular beam epitaxy. A Si-δ-doped $Al_{0.3}Ga_{0.7}As$ layer (sheet doping density $N_s=5 \times 10^{12} cm^{-2}$, spaced 5nm from the upper interface) of 220nm thickness was grown onto a semi-insulating GaAs substrate and subsequently covered with an undoped GaAs layer (90nm), an undoped $Al_{0.5}Ga_{0.5}As$ spacer layer (3nm) and a layer of Be-doped $Al_{0.5}Ga_{0.5}As$ (47nm, $N_a=8 \times 10^{18} cm^{-3}$). On top, a Be-doped GaAs capping layer was added. The layer structure was designed to accommodate two parallel conducting layers within the 90nm thick *i*-GaAs region: a hole gas at the upper interface and an electron gas at the lower interface. In the as-grown state, the electron gas is fully depleted such that the hole gas forms the only conducting layer. Removing the upper, *p*-type material by wet-etching lowers the electrostatic potential of the *i*-GaAs channel such that the previously depleted electron gas forms at the lower interface. Thus, a junction between the electron and hole gas can be produced at the step-edge, which manifests itself in a variation of the electrostatic potential laterally along the *i*-GaAs channel. This situation is shown in the upper part of Figure 1, where dark areas in the cross-section of the structure correspond to low values of the electrostatic potential. The potential distribution was obtained using a 2D-Poisson solver. Also shown in the lower part of Figure 1 are simulated valence and conduction band energies along two lines in the *z*-direction. The following sections will investigate this cross-section with particular attention to the undoped GaAs region.

A through-the-lens detector in a Schottky Field Emission Gun SEM (FEI S-FEG XL30) was employed to collect SE images. The sample was cleaved directly prior to loading in the vacuum chamber (base pressure $3.0 \times 10^{-6}$ mbar in an oil-free vacuum system). The cross-section was aligned perpendicular to the electron beam at a working distance of 5.5 mm and a primary beam energy of 1keV. An extractor voltage of 6V was used in order to optimise the contrast [12].

A cross-section SE-image of the tilted device is shown in Figure 2. To the left of the step-edge the acceptor-doped AlGaAs layer was removed. Since a highly selective etch (HF 48%) was used the 5nm thick GaAs capping layer can still be seen covering the step edge. As expected, the *p*-AlGaAs region shows a strong SE signal, i.e. it is relatively bright. However, also the *i*-GaAs channel underneath the *p*-AlGaAs results in a large SE signal, and cannot be distinguished from *p*-AlGaAs, even though it is undoped. Contrast variations can be seen within the *i*-GaAs layer parallel to the interface. In this case, modulation of the escape barrier height $E_b$ in the *i*-GaAs region is caused by space charges located outside the *i*-GaAs channel, rather than by a local doping distribution.

In order to show that the observed contrast is consistent with the theoretical potential distribution, the simulated electric potential $\varphi(x, z)$ is compared with the SE emission intensity from an un-tilted sample in Figure 3. On the left, the SE micrograph in (a) and a 2D grey-scale plot of simulated $\varphi(x, z)$–values in (b) are shown. The similarity between these two plots is due to the relationship between the electrostatic potential $\varphi$ and the escape barrier height $E_b$. Both plots show that in the *i*-GaAs channel the gradient of $\varphi$ parallel to the interface becomes stronger with decreasing distance to the upper interface. This is because the parallel component of the field at the step-edge is generated by ionized acceptors from the upper AlGaAs layer and surface charges where this layer was removed.

The variation of the simulated potential $\varphi$ inside and along the *i*-GaAs channel is shown in the diagram of Figure 3 (c) (thick line). Underneath the region where the *p*-AlGaAs was removed ($x = 0$ to $0.4$ µm) the electrostatic potential is lower than in the as-grown state, where the *p*-AlGaAs was not etched ($x = 0.4$ to $0.8$ µm). This agrees well with the experimental variation of the SE intensity (thin line) along the line corresponding to the location that is indicated by the arrows to the right of the 2D plots. The lateral junction width has been obtained from this diagram and is about 30nm. This demonstrates not only the sharp lateral potential variation which can be produced in this way, but also the capability of the SEM to provide information on potential differences on a nanometre scale.

In contrast to conventional *p-n* junctions, this field is generated by ionized dopants outside the region of free charge carriers. This might lead to a dependence of the surface-charge layer on an applied external bias different from an ordinary *p-n* junction [13]. The SEM could therefore provide a useful tool to explain the negative differential resistance and bistability observed in the electrical characteristic of this lateral junction [14].

In summary, the cross-section of an undoped GaAs layer, into which an electron-hole-junction is embedded, was imaged using SEM. The contrast is demonstrated to be due to the spatial variation of the electrostatic potential, caused by external space charges, rather than to conventional doping contrast. This interpretation was supported by a 2D Poisson simulation. Secondary electron imaging in the SEM can therefore be used to provide information on potential differences on the nanometer scale in both doped and undoped materials.

The authors thank FEI for financial support. B. Kaestner thanks D. A. Williams for his valuable input and wishes to acknowledge Prof. C. R. Stanley and Dr. J. L. Pearson for discussions about the wafer design and the Cavendish Laboratory for the provision of a J. J. Thomson studentship.

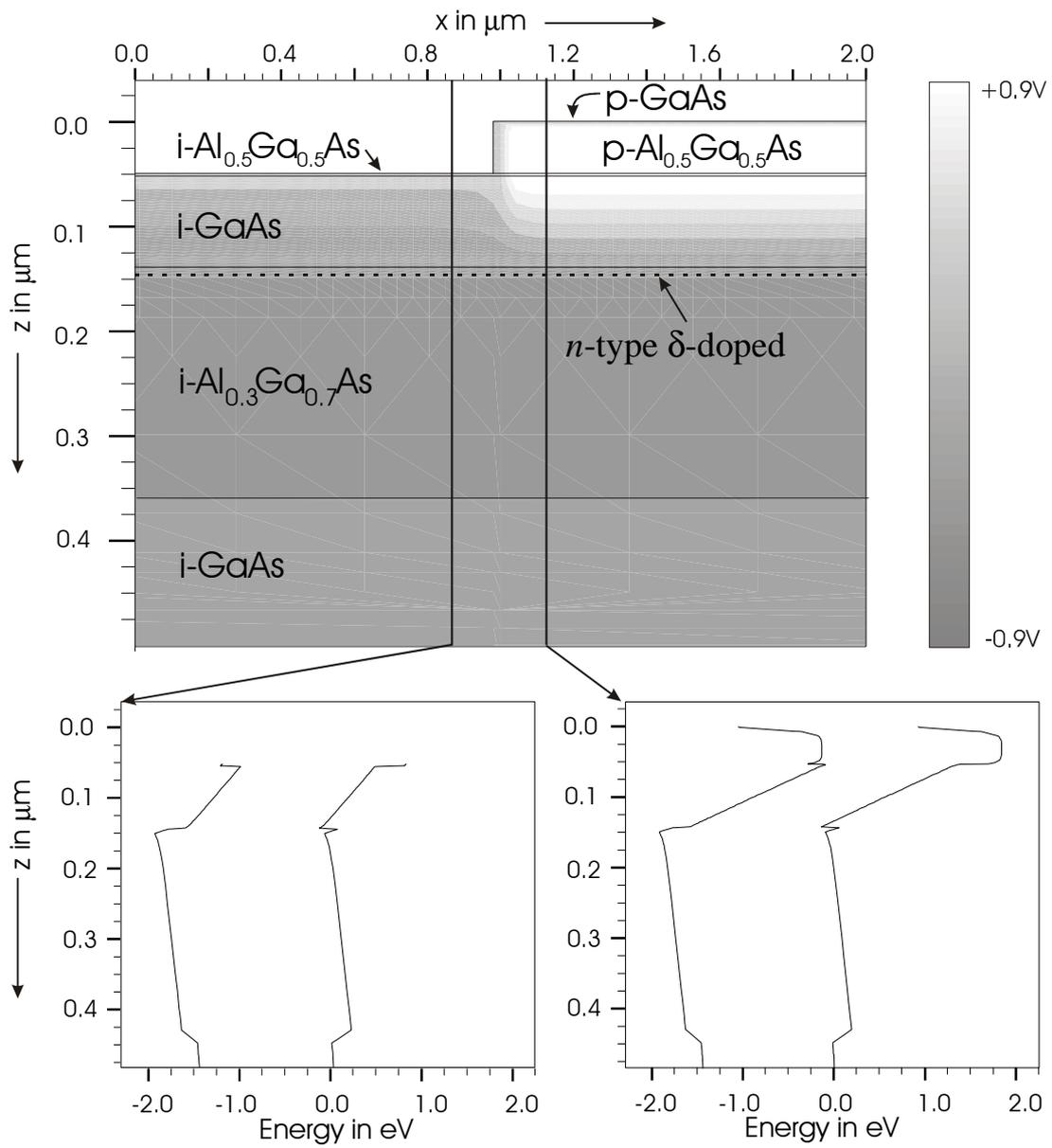

Figure 1

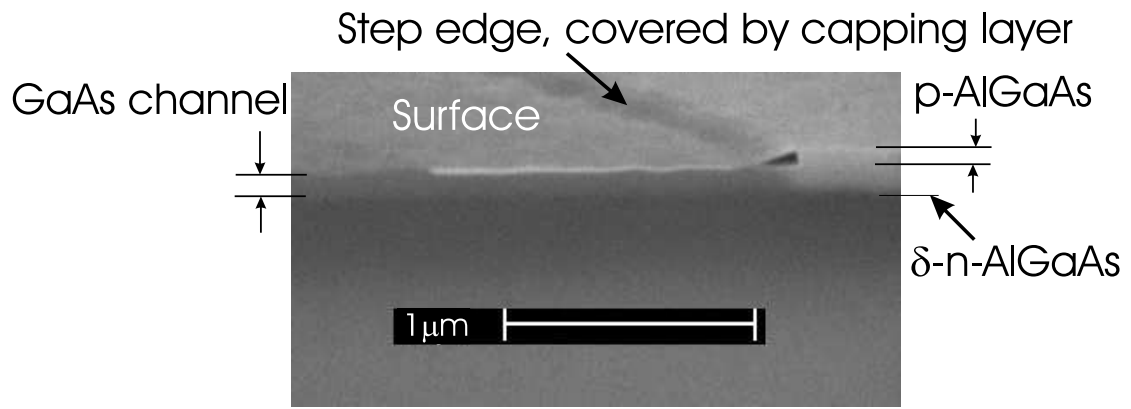

Figure 2

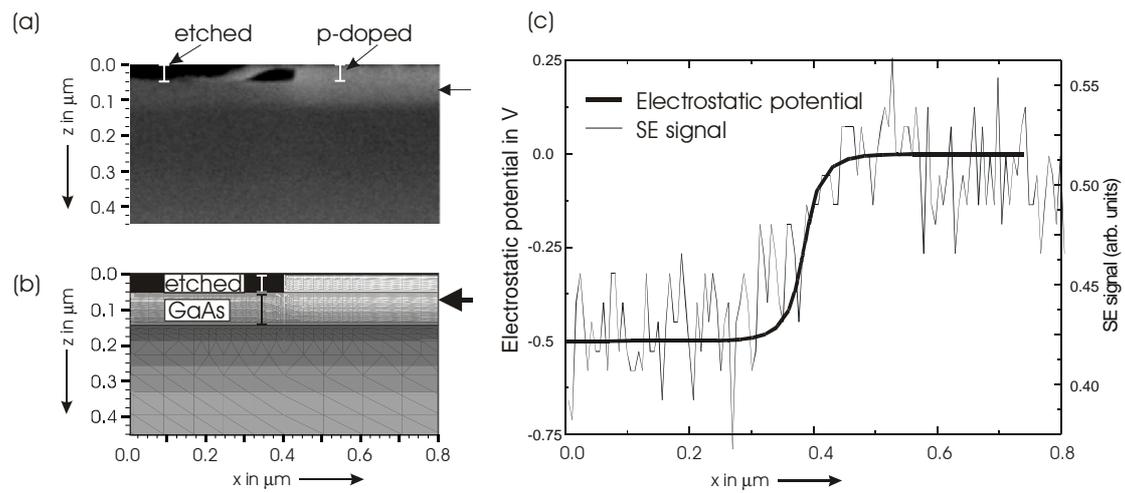

Figure 3

**List of Figures:**

**Figure 1**

Cross-section of the structure, showing the simulated potential distribution in grey-scale. It varies both laterally and vertically within the undoped GaAs channel. Simulated band diagrams along two lines in $z$-direction to either side of the step edge are shown at the bottom.

**Figure 2**

SEM image of tilted sample. The undoped GaAs channel (90nm in height) extends from the surface to the left of the step edge, where the 50nm $p$-AlGaAs layer was removed, down to the interface of the δ-$n$ doped AlGaAs region, which appears dark. Brightness variation from right to left can be seen within the undoped GaAs channel, similar to the simulated potential profile in Figure 1.

**Figure 3**

A secondary electron (SE) micrograph (a) is compared to the simulated electric potential $\varphi(x, z)$ (b) of the junction cross-section. Light areas indicate high $\varphi$-values. The diagram in part (c) compares the variation of simulated $\varphi$-values (thick line) along a cut line in $x$-direction to the corresponding SE signal (thin line). The $z$-position of the cut lines is indicated by the thick and thin arrows respectively, and lies in the undoped GaAs channel.